\documentclass[twocolumn,notitlepage]{revtex4-2}
\usepackage{multirow}
\usepackage{xcolor}
\usepackage{graphicx}
\usepackage{tikz-cd}
\usepackage{dcolumn}
\usepackage[normalem]{ulem}
\usepackage{amsmath}
\usepackage{amssymb}
\usepackage{mathtools}
\usepackage{hyperref}
\usepackage{cancel}
\allowdisplaybreaks

\newenvironment{remark}[1][Remark]{\begin{trivlist}
\item[\hskip \labelsep {\bfseries #1}]}{\end{trivlist}}

\newcommand{\qed}{\nobreak \ifvmode \relax \else
	\ifdim\lastskip<1.5em \hskip- \lastskip
	\hskip 0.5em plus0em minus0.5em \fi \nobreak
	\vrule height0.75em width0.5em depth0.25em\fi}

\begin{document}

\title{Causal dynamics of null horizons under linear perturbations}

\author{Peter K S \surname{Dunsby}}
\email{peter.dunsby@uct.ac.za}
\affiliation{Cosmology and Gravity Group, Department of Mathematics and Applied Mathematics, University of Cape Town, Rondebosch 7701, Cape Town, South Africa}
\affiliation{South African Astronomical Observatory, Observatory 7925, Cape Town, South Africa}
\affiliation{Centre for Space Research, North-West University, Potchefstroom 2520, South Africa}

\author{Seoktae \surname{Koh}}
\email{kundol.koh@jejunu.ac.kr}
\affiliation{Department of Science Education, Jeju National University, Jeju, 63243, South Korea}

\author{Abbas M \surname{Sherif}}
\email{abbasmsherif25@gmail.com}
\affiliation{Department of Science Education, Jeju National University, Jeju, 63243, South Korea}

\begin{abstract}
 We study the causal dynamics of an embedded null horizon foliated by marginally outer trapped surfaces (MOTS) for a locally rotationally symmetric background spacetime subjected to linear perturbations. We introduce a simple procedure which characterizes the transition of the causal character of the null horizon. We apply our characterization scheme to non-dissipative perturbations of the Schwarzschild and spatially homogeneous backgrounds. For the latter, a linear equation of state was imposed. Assuming a harmonic decomposition of the linearized field equations, we clarify the variables of a formal solution to the linearized system that determine how the null horizon evolves. For both classes of backgrounds, the shear and vorticity 2-vectors are essential to the characterization, and their roles are made precise. Finally, we discuss aspects of the relationship between the characterizing conditions. Various properties related to the self-adjointness of the MOTS stability operator are extensively discussed. 
\end{abstract}

\maketitle

\section{Introduction}


\noindent\textit{Background and motivation}: The discovery of supermassive black holes as well as the detection of gravitational waves emitted by inspiralling black holes have ignited renewed interest in black hole physics. A key concept in understanding the dynamics of black holes is the notion of a trapped surface, those on which both ingoing and outgoing null rays are converging, introduced in the proof of the 1965 singularity theorem by Roger Penrose \cite{rp1}.

Marginally outer trapped surfaces (MOTS)  are the marginal case of trapped surfaces, closed surfaces on which the outgoing null ray is neither converging nor diverging. Slight iterations to the definition of MOTS can be found in the literature from Wald's marginally trapped surface \cite{wd1} (note the ``outer'' here has been dropped) which rather imposes that both null rays are non-diverging, to Hayward's marginal surface \cite{hay1} which is in fact what we call a MOTS here, but in Hayward's case the choice is which of the null ray that is neither converging nor diverging is a not fixed. (See the references \cite{ib1,ib3} for excellent reviews and additional details. The notion of a MOTS will also be rigorously defined in Section \ref{sec3}).

MOTS foliate 3-dimensional hypersurfaces known as \textit{marginally outer trapped tubes} (MOTTs) \cite{ash2}, and where the causal character is fixed at all points of a MOTT it will be referred to by the common name \textit{horizon}, which under certain conditions bound trapped regions containing trapped surfaces. Dynamical and null horizons are well known cases of horizons (see for example \cite{ash1}).  The timelike MOTTs are usually referred to simply as \textit{membranes} as they, by definition, cannot enclose trapped surfaces.

MOTS have also been fundamental to understanding the formation of black holes due to gravitational collapse, as well as their dynamics and evolution. MOTS admit a notion of stability \cite{and1,and2} (consequently leading to a notion of stability of black holes). This is a way to determine whether one can deform a MOTS $\mathcal{S}$ to another $\mathcal{S'}$ along a unit direction normal to $\mathcal{S}$ that points along the slice in which $\mathcal{S}$ is embedded. This stability condition is given in terms of conditions on the principal eigenvalue of a second order linear elliptic operator. It has been shown that under physically reasonable conditions, strict stability, where the principal eigenvalue is positive, a MOTS will evolve to foliate a DH \cite{and1} (the same conclusion follows in the case that the operator has no zero eigenvalue).

Given a horizon embedded in a background spacetime which is subjected to a perturbation, one expects the perturbation to in general affect the dynamics, and hence causal character of the horizon. Perturbation of horizons have been examined variously in different contexts using different approaches. For example, in \cite{km1} linear perturbations of a null horizon by a gravitational field was studied and it was established that such perturbations do not affect the causal character of the horizon. This feature of the invariance of the causal character, as have already been mentioned, will not be the case in general, and this was neatly discussed in \cite{ash20}, where generic first and higher order perturbations of non-expanding horizons were considered.

A more explicit example was considered by Pilkington \textit{et al.} in \cite{pil1}. Embedding a stationary metric in the Weyl solution is a way of smoothly distorting stationary blackholes (see for example \cite{gero1}). The Weyl potentials tune the behavior of horizons and are referred to as \textit{distortion potentials}. The authors demonstrated that while for small distortions the horizon is apparently `well behaved', for large distortions, even the MOTS structure of the horizon is not assured, let alone the causal character.

\ \\
\noindent\textit{Objectives}: Our aim here is to introduce an approach which characterizes the evolution of MOTS in a linearly perturbed locally rotationally symmetric (LRS) background spacetime \cite{el1,el2}. More precisely we consider the following problem: How does a null horizon foliated by MOTS, embedded in a LRS background solution, evolve when the background is subjected to linear perturbations? 

We will employ the 1+1+2 covariant formalism \cite{cc1,cc2}, which has been used in several works to study MOTS and their evolution in background LRS spacetimes \cite{el3,as1,as2,as3}. After formulating the characterization scheme, we will adapt our approach to cases of some well known background solutions. Specifically, we consider the null event horizon in the Schwarzschild background and null horizons in hypersurface orthogonal and spatially homogeneous background solutions. As will be seen, at least in these examples to which our scheme will be applied, the shear and vorticity 2-vectors appear to consistently play the pivotal roles in determining the causal behavior of the null horizons under perturbations. (The roles of these 2-vectors will be made more precise in the text).

We also aim to consider aspects of the self-adjointness of the stability operator. As it will be seen, the 1-form that characterizes the self-adjointness of the operator is defined in terms of the shear and vorticity 2-vectors as well, and so it is quite natural to consider this in relation to the evolution of the horizons.

\ \\
\noindent\textit{Structure}: Section \ref{sec2} of the paper gives a quick overview of the covariant formalism that is used, and the procedure for linearizing a LRS background solution, which provides the mapping between a background solution and a perturbed one, is outlined. The process to harmonically decompose first order scalars, 2-vectors and 2-tensor, in this formalism is briefly reviewed following the standard literature. In Section \ref{sec3}, we introduce an approach to characterize MOTS evolution in the perturbed solution. 

We start by briefly introducing the notion of a MOTS, which is well proliferated in the literature but nonetheless necessary for the flow of the paper. We then go on to introduce the characterization scheme where it becomes transparent how the required gauge choice forces us to consider a null horizon in the background. That is, our scheme does not work for a non-null horizon. Section \ref{sec4} applies the characterization scheme to static and spatially homogeneous LRS backgrounds. In both cases the evolution of the null horizon is characterized, with the case of the latter an equations of state being imposed and a partial characterization provided.  In Section \ref{sec5}, we discuss some simple implications of the results of Section \ref{sec4} for the self-adjointness of the MOTS stability. Several additional properties of the MOTS and aspects of the stability operator are discussed. We conclude with discussion of results in Section \ref{sec6} and discuss potential avenues for future research.


\section{Preliminaries}\label{sec2}


In this section we briefly introduce the 1+1+2 covariant formalism. We then describe the general prescription for the perturbative scheme that have been used extensively over the last few years, to carry out perturbations of background LRS solutions.

\subsection{Covariant decomposition}

There is by now a well established trove of literature on the 1+1+2 covariant formalism, with several references providing details of the steps in the decomposition. A reader interested in more details is referred to the following standard texts \cite{cc1,cc2} (Also see the references \cite{sem1,sem2,chev1} for some recent corrections to the full set of equations). As such, the introduction here would be very brief, only necessary enough for the work carried out in this paper.

In the well known and powerful 1+3 approach, a unit tangent direction - call this \(u^a\) - along which timelike observers flow, is what threads the spacetime. The field equations are decomposed along \(u^a\) plus some constraint equations. If there exists some unit spacelike vector \(n^a\) obeying \(u_an^a=0\), the 3-space from the previous split can be further decomposed along this direction as a \(1+2\) product manifold. This, in addition to the \textit{evolution} equations along \(u^a\), introduces a set of \textit{propagation} equations along \(n^a\), as well as some constraint equations. The splitting naturally introduces new derivative along \(n^a\) and on the 2-space (generally referred to as the \textit{sheet}):

\begin{itemize}
\item \textit{Along \(u^a\)}: \\ \(\dot{\psi}^{a\cdots b}_{\ \ \ \ c\cdots d}=u^f\nabla_f\psi^{a\cdots b}_{\ \ \ \ c\cdots d}\);

\item \textit{Along \(n^a\)}: \\ \(\hat{\psi}^{a\cdots b}_{\ \ \ \ c\cdots d}=n^f\nabla_f\psi^{a\cdots b}_{\ \ \ \ c\cdots d}\);

\item \textit{Along the sheet}:\\ \(\delta_f\psi^{a\cdots b}_{\ \ \ \ c\cdots d}=N^{\bar c}_{\ c}\cdots N^{\bar d}_{\ d}N^a_{\ \bar a}\cdots N^b_{\ \bar b}N^e_{\ f}\nabla_e\psi^{\bar a\cdots \bar b}_{\ \ \ \ \bar c\cdots \bar d}\),
\end{itemize}
for any tensor \(\psi^{a\cdots b}_{\ \ \ \ c\cdots d}\), where \(N^{ab}=g^{ab}+u^au^b-n^an^b\) projects vectors and tensors to the sheet (it is the metric induced from the splitting, on the sheet).

A spacetime vector \(\psi^a\) can accordingly be decomposed as \(\psi^a=\Psi u^a+\bar\Psi n^a+\Psi^a\) (\(\Psi^a\) is the sheet component of \(\psi^a\)), and a fully projected tensor \(^3\psi_{ab}\) on the 3-space decomposes as

\begin{eqnarray}\label{dec}
^3\psi_{ab}=\Psi\left(n_an_b-\frac{1}{2}N_{ab}\right)+\Psi_{(a}n_{b)}+\Psi_{ab},
\end{eqnarray}
with $\Psi=-^3\psi_{ab}N^{ab}$, $\Psi_a=N_a^{\ b}\  ^3\psi_{bc}n^c$ and

\begin{align}
\Psi_{ab}=\left(N_{(a}^{\ c}N_{b)}^{\ d}-\frac{1}{2}N_{ab}N^{cd}\right) \ ^3\psi_{cd},
\end{align}
is the symmetric, fully projected and trace-free part of \(^3\psi_{ab}\). The above can be seen as the 1+1+2 analogue of the 1+3 orthogonally projected symmetric trace-free part of a 4-tensor:

\begin{align}
\psi^{\langle ab\rangle}=\left(h^{(a}_{\ c}h^{b)}_{\ d}-\frac{1}{3}h^{ab}h_{cd}\right)\psi^{cd},
\end{align} 
where $h_{ab}=g_{ab}+u_au_b$ which introduces the derivative operator

\begin{align}
D_f\psi^{a\cdots b}_{\ \ \ \ c\cdots d}=h^{\bar c}_{\ c}\cdots h^{\bar d}_{\ d}h^a_{\ \bar a}\cdots h^b_{\ \bar b}h^e_{\ f}\nabla_e\psi^{\bar a\cdots \bar b}_{\ \ \ \ \bar c\cdots \bar d}.\nonumber
\end{align}

So, for example the shear, electric and magnetic Weyl 3-tensors decompose respectively as

\begin{align}
\sigma_{ab}&=\sigma\left(n_an_b-\frac{1}{2}N_{ab}\right)+\Sigma_{(a}n_{b)}+\Sigma_{ab},\\
E_{ab}&=\mathcal{E}\left(n_an_b-\frac{1}{2}N_{ab}\right)+\mathcal{E}_{(a}n_{b)}+\mathcal{E}_{ab},\\
H_{ab}&=\mathcal{H}\left(n_an_b-\frac{1}{2}N_{ab}\right)+\mathcal{H}_{(a}n_{b)}+\mathcal{H}_{ab}.
\end{align}

Similarly, gradients of scalars decompose along these directions:

\begin{eqnarray}\label{decn}
\nabla_a\psi=-\dot{\psi}u_a+\hat{\psi}n_a+\delta_a\psi.
\end{eqnarray}

The energy momentum tensor takes the covariant form

\begin{eqnarray*}
T_{ab}=\rho u_au_b+ph_{ab}+2q_{(a}u_{b)} +\pi_{ab}.
\end{eqnarray*}
Here, \(\rho=T_{ab}u^au^b\) is the (local) energy density, \(3p=T_{ab}h^{ab}\) is the (isotropic) pressure, \(q_a=T_{bc}h^b_{\ a}u^c\) is the heat flux vector ($q_a=Qn_a+Q_a$), and \(\pi_{ab}\), which can be decomposed using \eqref{dec} (with \(\Pi=T_{ab}n^an^b-p\)), captures the degree of anisotropy.

From the Einstein field equations with cosmological constant

\begin{align}
R_{ab}-\frac{1}{2}Rg_{ab}+\Lambda g_{ab}=T_{ab},\label{efe1}
\end{align}
where $R_{ab}$ and $R$ are respectively the spacetime Ricci tensor and scalar curvature, one can write down the Ricci tensor of the spacetime in the covariant form

\begin{align}
R_{ab}&=g_1u_au_b+g_2n_an_b+2\left(Qu_{(a}+\Pi_{(a}\right)n_{b)}\nonumber\\
&+\left(g_2-\frac{3}{2}\Pi\right)N_{ab}+2Q_{(a}u_{b)}+\Pi_{ab},\nonumber
\end{align}
where

\begin{align}
g_1=\frac{1}{2}\left(\rho+3p-2\Lambda\right)\quad\mbox{and}\quad g_2=\frac{1}{2}\left(\rho-p+2\Lambda+2\Pi\right).\nonumber
\end{align}

The covariant derivatives of the unit vector fields can be written down in terms of the kinematic (geometrical) variables:

\begin{subequations}
\begin{align}
\nabla_au_b&=-u_a\left(\mathcal{A}e_b+\mathcal{A}_b\right)+\left(\frac{1}{3}\theta+\sigma\right)n_an_b\nonumber\\
&+\frac{1}{2}\left(\frac{2}{3}\theta-\sigma\right)N_{ab}+\Omega\varepsilon_{ab}+\Sigma_{ab}\nonumber\\
&+2\left(n_{(a}\Sigma_{b)}+n_{[a}\varepsilon_{b]c}\Omega^c\right),\label{cd1}\\
\nabla_an_b&=-u_a\left(\mathcal{A}u_b+\alpha_b\right)+\left(\frac{1}{3}\theta+\sigma\right)n_au_b\nonumber\\
&+\frac{1}{2}\phi N_{ab}+\xi\varepsilon_{ab}+\zeta_{ab}+n_aa_b\nonumber\\
&+\left(\Sigma_a-\varepsilon_{ac}\Omega^c\right)u_b.\label{cd2}
\end{align}
\end{subequations}
\(\mathcal{A}=n_a\dot{u}^a\) is acceleration, \(\theta=h^{ab}\bar D_au_b\) is expansion of \(u^a\), \(\phi\) is expansion of \(n^a\) (called sheet expansion), \(\sigma=\sigma_{ab}n^an^b\) is the shear, \(2\Omega=\varepsilon^{ab}\nabla_au_b\) and \(2\xi=\varepsilon^{ab}\nabla_an_b\) are the respective twists of \(u^a\) and \(n^a\) (\(\Omega\) is usually referred to as vorticity or rotation, and \(\varepsilon_{ab}\) is the 2-dimensional alternating tensor); $\mathcal{A}^a$ is the sheet component of \(\dot{u}_a\); \(\alpha_a\) is the sheet component of \(\dot{n}_a\); \(a_a=\hat{n}_a\) is the acceleration of \(n^a\) and; \(\zeta_{ab}=\delta_{\{a} n_{b\}}\) is the shear of \(n^a\).

The field equations then takes a covariant form as a set of evolution and propagation equations of the covariant scalars, vectors and tensors, along with some constraint equations. These are obtained from the field equations themselves, and the Ricci and Bianchi identities for the unit fields $u^a$ and $n^a$. We do not need all of these equations here and will not enlist them. The few equations that will be needed for the current work will be introduced accordingly.

\subsection{Mapping the background and `true' spacetimes}


Perturbative schemes to study LRS cosmologies have been recently implemented using the 1+1+2 formalism. They have largely been applied to the LRS II class of spacetimes, with a recent consideration being a generalization which introduces dissipation of the fluid via the non-vanishing of the anisotropic stress on the background (see for example these recent works \cite{sem1,sem2}). 

In this section,  we begin by first introducing LRS spacetimes in context of the 1+1+2 formulation, and we briefly discuss the linearization procedure (more details can be found in the references \cite{cc1,cc2}), up to where it is relevant to this paper.

\textit{Locally rotationally symmetric (LRS)} spacetimes admit a continuous isotropy group at each point, as well as a preferred spatial direction (see the references \cite{el1,el2} for more details).

The symmetry of these spacetimes allows one to write the metric in local coordinates \((t,\mathcal{R},y,z)\) as (see \cite{el2} for example)

\begin{align}\label{fork}
ds^2&=-A_1^2dt^2+ A_2^2dr^2+ A_3^2dy^2\nonumber\\
&+\left(\left(DA_3\right)^2+\left(A_2h\right)^2-\left(A_1g\right)^2\right)dz^2\nonumber\\
&+2\left(A_1^2gdt-A_2^2hdr\right)dz,
\end{align}
where \(A_i=A_i\left(t,r\right)\), and \(g=g(y)\), \(h=h(y)\). \(D=D(y,k)\) where \(k\) is a constant and specified \(D\): For \(k=-1\), \(D=\sinh y\); for \(k=0\), \(D=y\) and; for \(k=1\), \(D=\sin y\). 

In the limiting case that \(g=0=h\), we recover the well studied LRS II class of spacetimes, which generalizes spherically symmetric solutions to the Einstein field equations. This class includes FLRW models, LTB spacetime model, Schwardschild solution, Oppenheimer--Snyder dust model, etc.

The covariant set

\begin{eqnarray}
\mathcal{D}:\equiv\lbrace{\mathcal{A},\theta,\sigma,\phi,\xi,\Omega,\rho,p,Q,\Pi,\mathcal{E},\mathcal{H}\rbrace},
\end{eqnarray}
specifies a LRS spacetime.

In general, the mapping between the background and perturbed spacetime generally depends on the problem of physical interest. And for the specific background under consideration, the linearization procedure may or may not be particularly cumbersome. Essentially, the problem reduces to determining which quantities vanish in the background and which do not. Here, we outline the procedure for the perturbation. (A detailed discussion on this is found in \cite{cc2}, formulated differently but unchanging underlying concept.)

\begin{itemize}

\item We first define the background spacetime: start with a background LRS spacetime \(M\), and let the collection of scalars 

\begin{eqnarray}
\mathcal{D}_1:\equiv\lbrace{\varphi^i\rbrace}_{i\in I},
\end{eqnarray}
for some indexing set \(I\), specify \(M\), i.e. \(\mathcal{D}_1\subset\mathcal{D}\). (Note that all 2-vector and 2-tensor quantities, those defined on the sheet, vanish in the background). 

\item Define the set

\begin{eqnarray}
\mathcal{D}_2:\equiv\mathcal{D}\setminus\mathcal{D}_1,
\end{eqnarray}
of those covariant scalars in \(\mathcal{D}\) that vanish in the background. 

\item Further introduce the set of 2-vectors and 2-tensors:

\begin{eqnarray}
\mathcal{D}_3:\equiv\lbrace{\Psi_a^j,\Psi_{ab}^w\rbrace}_{j\in J,w\in W},
\end{eqnarray}
with \(J\) and \(W\) being indexing sets, which also vanish in the background.

\item The full set of gauge invariant quantities is then

\begin{eqnarray}
\mathcal{D}_2\cup\mathcal{D}_3,
\end{eqnarray}
which follows from the Stewart-Walker Lemma \cite{stew1}. One can then use the Ricci and Bianchi identities for $u^a$ and the direction of anisotropy $n^a$  to obtain the linearized equations for the set $\bar D:\equiv\mathcal{D}_1\cup\mathcal{D}_2\cup\mathcal{D}_3$. 

The linearized system is generally not gauge invariant. The evolution and/or propagation equations of the background quantities may contain terms that do not vanish in the background. This is resolved by introducing $\delta$ gradients of the background quantities whose evolution and propagation equations replace those of the background scalars. 

\item We note that the `type' of perturbation being carried out will further simplify the collection $\mathcal{D}_2\cup\mathcal{D}_3$. For example, one may consider an irrotational or a shear-free perturbation, eliminating all shear and vorticity quantities.

There is also a degree of freedom in choosing the frame as $n^a$ can be freely chosen. This choice will depend on the particular physical consideration at hand. One common choice when dealing with black hole perturbation is to set the sheet component of the acceleration to zero \cite{cc1} which corresponds to a hovering observer. (At some point in this work we will make this choice of frame to transparently demonstrate our characterization scheme.) Other choices include the aligning of $n^a$ along the vorticity or the electric Weyl scalar. That is, one makes the respective frame choices $\Omega^a=0$ and $\mathcal{E}^a=0$. 

\item In general the linearized equations are quite messy and seldom tractible, analytically. This generally results from the presence of the angular `$\delta$' derivative. One approach is to decompose first order variables into harmonics, which converts the linearized equations into ordinary differential equations which in some cases can be solved exactly.

\end{itemize}

In this work we do not seek to solve the linearized system. Rather, we suppose that the harmonically decomposed linearized system can be solved exactly for an appropriate basis. Then, we specify the roles that certain components of the \textit{even} (\textit{electric}) vector harmonics  (to be defined shortly) play in the evolution of a null horizon embedded in a LRS background which is subject to a perturbation. 

\subsection{Harmonic decomposition}

We now present a quick overview of the harmonic decomposition procedure adapted to the formalism. Here we again follow the general setup in the references \cite{cc1,cc2}.

One starts by introducing dimensionless harmonic functions that satisfy
\begin{eqnarray}\label{harm1}
\delta^2 \mathcal{Q}=-\frac{\bar{k}^2}{r^2}\mathcal{Q},\quad\hat{\mathcal{Q}}=\dot{\mathcal{Q}}=0,\quad\quad\quad\quad\mbox{(\(0\leq\bar{k}^2\))}
\end{eqnarray}
on any LRS background, i.e. they are eigenfunctions of the sheet Laplacian. The function \(r\) is covariantly specified by the relations
\begin{eqnarray}\label{harm2}
\dot{r}=\frac{r}{2}\left(\frac{2}{3}\theta-\sigma\right),\quad\hat{r}=\frac{r}{2}\phi,\quad\delta_ar=0.
\end{eqnarray}

One may decompose any first order scalar \(\Psi\) as
\begin{eqnarray}\label{bas1}
\psi^{(1)}=\sum \psi_S^{(\bar{k})}\mathcal{Q}^{(\bar{k})},
\end{eqnarray}
usually written simply as $\psi^{(1)}=\psi_{S}\mathcal{Q}$, with the sum implicit over \(\bar{k}\), and the \(S\) subscript signalling a scalar decomposition.

The 2-vectors are also expanded in harmonics via the following defined basis

\begin{subequations}
\begin{align}
\mathcal{Q}^{\bar{k}}_a&=r\delta_a\mathcal{Q}^{\bar{k}}\implies N^b_{\ a}\hat{\mathcal{Q}}_b=0=N^b_{\ a}\dot{\mathcal{Q}}_b,\nonumber\\
\delta^2\mathcal{Q}_a&=\frac{1}{r^2}\left(1-\bar{k}^2\right)\mathcal{Q}_a;\label{vh1}\\
\bar{\mathcal{Q}}^{\bar{k}}_a&=r\varepsilon_{ab}\delta^b\bar{\mathcal{Q}}^{\bar{k}}\implies N^b_{\ a}\hat{\bar{\mathcal{Q}}}_b=0=N^b_{\ a}\dot{\bar{\mathcal{Q}}}_b,\nonumber\\
\delta^2\bar{\mathcal{Q}}_a&=\frac{1}{r^2}\left(1-\bar{k}^2\right)\bar{\mathcal{Q}}_a\label{vh2},
\end{align}
\end{subequations}
respectively named \textit{even} (electric) and \textit{odd} (magnetic) vector harmonics. (For all that are to follow, there should be no confusion that is to arise if we drop the superscript \(\bar{k}\) in relevant expressions.) 

Notice that the harmonic basis vectors \(\mathcal{Q}_a\) and \(\bar{\mathcal{Q}}_a\) can be obtained from one another, up to sign reversal, by applying \(\varepsilon_{ab}\): \(\bar{\mathcal{Q}}_a=\varepsilon_{ab}\mathcal{Q}^b\iff\mathcal{Q}_a=-\varepsilon_{ab}\bar{\mathcal{Q}}^b\). For this reason, \(\varepsilon_{ab}\) is understood as a parity operator, in the sense of the harmonics introduced here. A quick check also shows that

\begin{subequations}
\begin{align}
\delta_a\bar{\mathcal{Q}}^a&=0=\varepsilon_{ab}\delta^a\mathcal{Q}^b,\nonumber\\
\delta_a\mathcal{Q}^a&=-\frac{\bar{k}^2}{r}\mathcal{Q}=-\varepsilon_{ab}\delta^a\bar{\mathcal{Q}}^b\nonumber.
\end{align}
\end{subequations}
Since for each \(\bar{k}\), \(\mathcal{Q}_a\bar{\mathcal{Q}}^a=0\), first order vectors (\(2\)-vectors) can always be decomposed as

\begin{eqnarray}\label{basn}
\Psi_a=\sum \Psi_V^{(\bar{k})}\mathcal{Q}_a+\bar{\Psi}_V^{(\bar{k})}\bar{\mathcal{Q}}_a=\Psi_V\mathcal{Q}_a+\bar{\Psi}_V\bar{\mathcal{Q}}_a,
\end{eqnarray}
again, where the sum over \(\bar{k}\) is implicit, with the \(V\) signalling a vector decomposition. We note the following important point: for any first order 2-vector \(\Psi^a\), a quick check gives its sheet divergence as

\begin{eqnarray}\label{div1}
\delta_a\Psi^a=-\frac{\bar{k}^2}{r}\Psi_VQ.
\end{eqnarray}
Thus, \(\Psi^a\) is gradient if and only if \(\Psi_V\) vanishes.

For a \(2\)-tensor \(\Psi_{ab}\), the even and odd parity tensor harmonics are defined respectively as

\begin{subequations}
\begin{align}
\mathcal{Q}_{ab}&=r^2\delta_{\{a}\delta_{b\}}\mathcal{Q}\implies\hat{\mathcal{Q}}_{ab}=0=\dot{\mathcal{Q}}_{ab},\label{th1}\\
\bar{\mathcal{Q}}_{ab}&=r^2\varepsilon_{c\{a}\delta^c\delta_{b\}}\mathcal{Q}\implies\hat{\bar{\mathcal{Q}}}_{ab}=0=\dot{\bar{\mathcal{Q}}}_{ab}\label{th2}.
\end{align}
\end{subequations}
Again, we see that \(\varepsilon_{ab}\) is a parity operator here: \(\bar{\mathcal{Q}}_{ab}=\varepsilon_{c\{a}\mathcal{Q}_{b\}}^{\ \ c}\iff\mathcal{Q}_{ab}=-\varepsilon_{c\{a}\bar{\mathcal{Q}}_{b\}}^{\ \ c}\). Indeed, for each \(\bar{k}\) \(\mathcal{Q}_{ab}\bar{\mathcal{Q}}^{ab}=0\) so that any \(\Psi_{ab}\) can be decomposed as

\begin{eqnarray*}
\Psi_{ab}=\Psi_T\mathcal{Q}_{ab}+\bar{\Psi}_T\bar{\mathcal{Q}}_{ab},
\end{eqnarray*}
with the \(T\) signalling a tensor decomposition.


\section{Characterizing MOTS evolution}\label{sec3}


We introduce a characterization scheme for the evolution of MOTS orthogonal to $u^a$ and $n^a$. We are interested in how MOTS in constant time slices in linearized LRS background geometries evolve. We begin with a quick definition of MOTS, after which we introduce the characterization. As will be seen, the particular problem we are trying to address, or more precisely, how our objective is formulated, is imposed by the required gauge invariance of the evolution characterizing function. 

\subsection{MOTS}

For the surfaces under consideration, the following must hold in order for `2-space' to be a genuine surface rather than simply a collection of tangent planes (see the reference \cite{cc2} for discussion):

\begin{eqnarray}
\xi=\Omega=a^a&=0;\nonumber\\
\Sigma^a+\varepsilon^{ab}\Omega_b-\alpha^a&=0.\label{sf1}
\end{eqnarray}
(The left hand side of the second line is known as the Greenberg vector.) In particular, these conditions ensure that the Lie bracket of \(u^a\) and \(n^a\) vanishes, i.e. \(u^a\) and \(n^a\) are surface forming, and that the operator \(\delta_a\) is a `true' covariant derivative for the surface \cite{cc2}.

Consider a 2-surface \(S\) in spacelike constant \(t\) slices, with unit spacelike normal \(n^a\). Outgoing and ingoing null normals are respectively

\begin{eqnarray*}
k^a=u^a+n^a;\quad l^a=\frac{1}{2}\left(u^a-n^a\right),
\end{eqnarray*}
normalized so that \(l_ak^a=-1\). There is a degree of freedom to rescale the null vectors as \(k^a\rightarrow fk^a, l^a\rightarrow f^{-1}l^a\), for \(f>0\). But as will be noted shortly, the MOTS character relevant here is invariant under such scaling. 

The induced metric on \(S\) is 

\begin{eqnarray}
\mathcal{F}_{ab}=g_{ab}+2k_{(a}l_{b)},
\end{eqnarray}
and the ingoing and outgoing expansions are, respectively,

\begin{align}
\chi=\mathcal{F}^{ab}\nabla_ak_b;\qquad\bar\chi=\mathcal{F}^{ab}\nabla_al_b.\nonumber
\end{align}

(For our current purposes, $\mathcal{F}_{ab}$ here is just $N_{ab}$.) By MOTS, we mean that on \(S\), one has \(\chi=0\). In the case that the sign of \(\bar\chi\) is constrained as \(\bar\chi<0\) on \(S\), MOTS is abbreviated to MTS. (A MTS is generally considered to be associated to horizons that enclose black holes in dynamical spacetimes.) One sees that, under the rescaling, the expansion becomes \(\chi_f=f\chi\) (similarly, \(\bar{\chi}_f=f^{-1}\bar{\chi}\)). Thus, it is clear that any sign conditions on \(\chi\) and \(\bar{\chi}\) is seen to be invariant under the freedom to scale the null vectors.

\subsection{MOTS evolution}

Because of the MOTS condition \(\chi=0\), we know that the gradient \(\nabla_a\chi\) is orthogonal to the MOTS. Hence the norm, which we notate by \(z\), specifies the causal character, pointwise:

\begin{align}
z>0\quad\mbox{(Timelike)};\quad z<0\quad\mbox{(Spacelike)};\ z=0\quad\mbox{(Null)}.\nonumber
\end{align}

Let us decompose the expansion \(\chi\) into zeroth (background) and first order scalars:

\begin{eqnarray*}
\chi=\chi_0+\chi_1,
\end{eqnarray*}
with the subscripts \(\ast_0\) and \(\ast_1\) denoting, respectively, zeroth and first orders. (In general this will be the case as this is a linear combination of the covariant scalars, some of which may or may not vanish in the background.) We expand the norm \(z\) as

\begin{align}\label{norm}
z&=\left(-\dot{\chi}_0^2+\hat{\chi}_0^2\right)+\left(-\dot{\chi}_1^2+\hat{\chi}_1^2\right)+2\left(-\dot{\chi}_0\dot{\chi}_1+\hat{\chi}_0\hat{\chi}_1\right)\nonumber\\
&+\left(\delta_a\chi_0\delta^a\chi_0+\delta_a\chi_1\delta^a\chi_1+2\delta_a\chi_0\delta^a\chi_1\right).
\end{align}
As we neglect the products of first order quantities, the second and fourth parenthesized terms of \eqref{norm} vanish and we have

\begin{eqnarray}\label{norm1}
z=\left(-\dot{\chi}_0^2+\hat{\chi}_0^2\right)+2\left(-\dot{\chi}_0\dot{\chi}_1+\hat{\chi}_0\hat{\chi}_1\right).
\end{eqnarray}

Indeed, it is clear that depending on the particular background solution one begins with, the above expression will simplify as will later be observed. 

Of course there will be some cases where even extending to the nonlinear regime, i.e. in cases of nonlinear perturbations, the characterization can proceed using the form of $z$ \eqref{norm1}, for example when $z$ is optical, i.e. when $\delta_az\delta^az$ vanishes on the MOTS. Let us consider some cases where this is true for a particular frame choice or a class of backgrounds: The commutation relations between the `$\delta_a$', and the dot and hat derivatives, acting on a scalar $\psi$, are given respectively as (we set $\Omega=\xi=a_a=0$)

\begin{align}
\delta_a\dot{\psi}-N^{\ b}_a\left(\delta_b\psi\right)^{\cdot}&=-\mathcal{A}_a\dot{\psi}+\left(\Sigma_a-\varepsilon_{ab}\Omega^b+\alpha_a\right)\hat{\psi}\nonumber\\
&+\frac{1}{2}\left(\frac{2}{3}\theta-\sigma\right)\delta_a\psi,\label{coom1}\\
\delta_a\hat{\psi}-N^{\ b}_a\widehat{\left(\delta_b\psi\right)}&=-\left(\Sigma_a-\varepsilon_{ab}\Omega^b\right)\dot{\psi}+\frac{1}{2}\phi\delta_a\psi.\label{coom2}
\end{align}

Now notice that the function $r$, introduced in the harmonic decomposition obeys

\begin{align}
\dot{r}+\hat{r}=\frac{r}{2}\chi.\nonumber
\end{align}
And upon using the equations \eqref{coom1} and \eqref{coom2} with $\psi=r$ we find that

\begin{align}
\delta_a\chi=\left(\mathcal{A}_a+\alpha_a\right)\phi.\label{coom3}
\end{align}
So, a frame choice $\mathcal{A}_a=\alpha_a=0$, $\delta_a\chi=0$. Also, on any background with a vanishing sheet expansion, it is true that $\delta_a\chi=0$ on the MOTS. Thus, analyzing MOTS in perturbed spatially homogeneous or static LRS backgrounds, the form of the function $z$ suffices even in nonlinear regimes, although the linearized equations is now more formidable. However, we will keep the forthcoming applications modest and restrict ourselves to linear perturbations.

As is easily seen from \eqref{norm1}, the scalar \(z\) will generally not be gauge invariant: Indeed, the second parenthesized terms do vanish in the background due to the presence of $\chi_1$ and derivatives thereof, but the first parenthesized ones do not. Characterization of the evolution of the MOTS in the perturbed spacetime obviously requires a gauge invariant $z$. In fact, $z$ is gauge invariant if and only if the combination of terms in the first parenthesis vanishes. It indeed follows that, were one to start with a null horizon $\bar{H}$, foliated by MOTS, in the background, $z$ is gauge invariant with respect to $\bar H$ since the vanishing of the sum in the first parenthesis characterizes the nullity of $\bar{H}$ in the first place. For this reason, we can more precisely frame what is being done in this work:\\
\ \\
\textit{Consider an exact LRS background spacetime $M$ and let $\bar{H}$ be a null horizon in $M$, foliated by MOTS. Let us suppose that $M$ is subject to a perturbation of linear order. How do we characterize the evolution of the horizon $\bar{H}$ under the perturbation?}\\
\ \\
Now we emphasize again that we are not interested in solving linearized system. It is assumed that an appropriate gauge (and possibly frame) choice has been made, so that one has a mapping between the background and the perturbed spacetimes. Independent of the gauge choice, Then, by simply imposing the vanishing condition

\begin{align}
-\dot{\chi}_0^2+\hat{\chi}_0^2=0,\nonumber
\end{align}
the function $z$ will characterize the evolution of a null horizon from the background. The gauge choice that maps the background and perturbed spacetimes of course determines the simplification of $z$ as it is constructed from elements of the linearized equations.


Now, to ensure a consistent sign of $z$ along a horizon foliated by the MOTS under consideration, one requires $\delta_az=0$. Explicitly, this is the requirement that

\begin{eqnarray}\label{norm2}
\dot{\chi}_0\delta_a\dot{\chi}-\hat{\chi}_0\delta_a\hat{\chi}=0.
\end{eqnarray}
Again, once a background is specified, the expression \eqref{norm2} simplifies as will be seen in Section \ref{sec4}. 

\begin{remark}[Remark 1.]
We point out that if one relaxes the $\delta_az=0$ condition, one allows for a range of possibilities, for example, horizons with different portions exhibiting varying causal characters. This is beyond the scope of the current work and will be deferred for a subsequent paper.
\end{remark}

Let us now compute the forms of the outgoing and ingoing null expansion scalars:

\begin{align}
\chi=\frac{2}{3}\theta-\sigma+\phi,\qquad\bar\chi=\frac{1}{2}\left(\frac{2}{3}\theta-\sigma-\phi\right).\label{chio}
\end{align}
Thus, a MOTS is specified by the covariant triple \((\theta,\sigma,\phi)\), and for any particular background, this splits into the background and first order parts. 

If our interest is in MOTS that are MTS, then the requirement that \(\bar\chi<0\) implies that necessarily \(\phi>0\). For our purposes here, however, we will not impose such restriction and will instead consider the more general `MOTS'.

We note that the form of $\chi$, first equation of \eqref{chio}, is form invariant with respect to the background and so the gauge problem is automatically fixed. This is a consequence of the particular class of MOTS we are considering for this work. For a different class of MOTS, this correspondence may be broken and some additional condition may have to be imposed to fix the gauge, in general. However, up to linear order for null horizons as is considered here (or more generally nonexpanding horizons) $\chi$ vanishes on the perturbed horizon as was established in \cite{ash20}.

\begin{remark}[Remark 2.]

Before proceeding, we point out that caution is to be exercised when considering the vanishing of the first parenthesized terms of \eqref{norm}. For example if one is to consider a static background, then obviously, $\dot{\chi}^2_0$ should vanish. However, $\hat{\chi}^2_0$ does not vanish on a null horizon in the background. So, the terms should be taken together. In fact, its easily seen that the first parenthesized terms of \eqref{norm} -- we denote this as \(F\) -- factors as

\begin{eqnarray}
F=-\mathcal{L}_k\chi\mathcal{L}_l\chi.
\end{eqnarray}
The vanishing condition in the background, assuming the null energy condition is then $\mathcal{L}_k\chi=0$. So, one has to take care of the function $F$, so that once a background is specified, $F$ vanishes on a null horizon in the background. 
\end{remark}

The evolution and propagation equations for \(\phi\) and \((2/3)\theta-\sigma\), up to linear order, are \cite{cc1}

\begin{subequations}
\begin{align}
\dot{\phi}&=\left(\frac{2}{3}\theta-\sigma\right)\left(\mathcal{A}-\frac{1}{2}\phi\right)+Q+\delta_a\alpha^a,\label{leb1}\\
\hat{\phi}&=-\frac{1}{2}\phi^2+\left(\frac{2}{3}\theta-\sigma\right)\left(\frac{1}{3}\theta+\sigma\right)-\mathcal{E}-\frac{1}{2}\Pi\nonumber\\
&-\frac{2}{3}\left(\rho+\Lambda\right)+\delta_aa^a,\label{leb2}\\
\frac{2}{3}\dot{\theta}-\dot{\sigma}&=\mathcal{A}\phi-\frac{1}{2}\left(\frac{2}{3}\theta-\sigma\right)^2+\mathcal{E}-\frac{1}{2}\Pi\nonumber\\
&-\frac{1}{3}\left(\rho+3p-2\Lambda\right)+\delta_a\mathcal{A}^a,\label{leb3}\\\
\frac{2}{3}\hat{\theta}-\hat{\sigma}&=\frac{3}{2}\phi\sigma+Q+\delta_a\left(\Sigma^a+\varepsilon^{ab}\Omega_b\right).\label{leb4}
\end{align}
\end{subequations}
Equation \eqref{leb4} is obtained by projecting along $n^a$ the shear divergence equation \cite{el4}

\begin{eqnarray*}
D^b\sigma_{ba}-\frac{2}{3}D_a\theta+\varepsilon_{abc}\left(D^b+2\dot{u}^b\right)\omega^c+q_a=0,
\end{eqnarray*}
where

\begin{eqnarray*}
\omega^a=\Omega n^a+\Omega^a,\quad q^a=Qn^a+Q^a.
\end{eqnarray*}
Equations \eqref{leb1}, \eqref{leb2}, and \eqref{leb3} are obtained, respectively, by applying $u^aN^{bc}$, $n^aN^{bc}$, and $-n^au^bu^c$ to the Ricci identities for $n^a$.

We can write down the evolution of the null expansion along the null rays as

\begin{align}
\mathcal{L}_k\chi&=\chi\left(\mathcal{A}-\frac{1}{2}\phi+\frac{3}{2}\sigma\right)-\left(\rho+p+\Pi\right)\nonumber\\
&+2Q+\delta_a\left(\alpha^a+\mathcal{A}^a+a^a+\Sigma^a+\varepsilon^{ab}\Omega_b\right),\label{gip1}\\
\mathcal{L}_l\chi&=-\frac{1}{2}\chi\left(\chi-\frac{1}{2}\phi+\frac{3}{2}\sigma\right)+\frac{1}{3}\left(\rho-3p\right)+2\mathcal{E}\nonumber\\
&+\delta_a\left(\alpha^a+\mathcal{A}a-a^a-\Sigma^a-\varepsilon^{ab}\Omega_b\right).\label{gip2}
\end{align}
Note that $\chi$ vanishes on the horizon and hence the above are simplified there. The scalar $F$ now expands, up to linear order on the horizon, to the form

\begin{align}\label{gip3}
F&=\left(\frac{1}{3}\left(\rho-3p\right)+2\mathcal{E}\right)\nonumber\\
&\times\left(\left(\rho+p+\Pi\right)-2Q-\delta_aZ^a\right)
\end{align}
where we have defined

\begin{align}
Z^a=\alpha^a+\mathcal{A}^a+a^a+\Sigma^a+\varepsilon^{ab}\Omega_b.
\end{align}

Depending on the particular class of background, and how one chooses the $\{u^a-n^a\}$ frame by appropriately fixing the $n^a$ direction, the scalar $F$ can be considerably simplified as will be seen in the next section. 

\subsubsection*{The future outer trapping condition}

A MOTS $\mathcal{S}$ on which the ingoing expansion $\bar\chi$ is strictly negative and $\mathcal{L}_l\chi<0$ is known as a \textit{future outer trapped surface} (FOTS). A horizon foliated by FOTS is a \textit{future outer trapping horizon} (FOTH). These were introduced by Hayward in \cite{hay1} and have since be extensively studied \cite{ib4,ib5}.) In the dynamical and isolated cases, respectively, assuming the null energy condition, one has that $\mathcal{L}_k\chi<0$ and $\mathcal{L}_k\chi=0$. In this case we have a dynamical FOTH and an isolated FOTH. Crucially, the FOTH condition immediately implies the presence of a black hole as it ensures the existence of trapped surfaces just to the inside of the horizon. Thus, the conditions characterizing a FOTH for the dynamical and isolated cases are $\bar\chi<0$, \eqref{norm2}, plus $\mathcal{L}_k\chi<0$ and $\mathcal{L}_k\chi=0$, respectively,

\begin{align}
z&<0,\dot{\chi}<0,\hat{\chi}<0,\label{zee1}\\
z&=0,\dot{\chi}<0,\hat{\chi}>0.\label{zee2}
\end{align}

Henceforth, when specifying to a FOTH the condition $\bar\chi<0$ will be assumed.

In the next section, we consider some backgrounds as particular examples, to apply the horizon characterization scheme developed in this section to horizons in the corresponding perturbed spacetime.


\section{Applications}\label{sec4}

In this section we consider some particular background LRS solutions. Specifically, we consider static and spatially homogeneous LRS backgrounds, with restriction to the class II. We will also briefly comment on the case of a background with both $t$ and $r$ dependence (see \textbf{Remark 3}). 

\subsection{Static background}

Let us take as an example a non-dissipative linear perturbation of the Schwarzschild background with metric (we set the mass to unity here, which does not, in any serious way, affect the calculations)

\begin{align}
ds^2=-\left(1-\frac{2}{r}\right)dt^2+\left(1-\frac{2}{r}\right)^{-1}dr^2+r^2d\varsigma^2,
\end{align}
where $d\varsigma^2$ is the metric on the unit 2-sphere. We know the above is time-symmetric, i.e. \(\theta=0=\sigma\). Hence, we have the zeroth and first orders decomposition of the null expansion $\chi$ as

\begin{eqnarray*}
\chi_0=\phi\qquad\mbox{and}\qquad\chi_1=\frac{2}{3}\theta-\sigma.
\end{eqnarray*}
Therefore, \(\dot{\chi}_0\) and \(\dot{\chi}_1\) are first order quantities so that \(\dot{\chi}_0^2=\dot{\chi}_0\dot{\chi}_1=0\). The only non-vanishing background covariant scalars are \(\mathcal{A},\phi,\mathcal{E}\), given in coordinate forms as

\begin{eqnarray*}
\mathcal{A}=\frac{1}{r^2}\left(\sqrt{1-\frac{2}{r}}\right)^{-1};\quad\phi=\frac{2}{r}\sqrt{1-\frac{2}{r}};\quad\mathcal{E}=-\frac{2}{r^3}.
\end{eqnarray*}

In this background the MOTS is located at $\phi=0\iff r=2$. Now, the norm \(z\) becomes

\begin{align}
z&=F+2\hat{\chi}_0\hat{\chi}_1\nonumber\\
&=-4\mathcal{E}\delta_a\left(Z^a+\Sigma^a+\varepsilon^{ab}\Omega_b\right).\label{norm3}
\end{align}

Additionally one also requires, from \eqref{norm2}, that (we have now set $a^a=0$ on the horizon as it should be)

\begin{eqnarray}\label{norm20}
0=\hat{\chi}_0\delta_a\hat{\chi}=\delta_a\mathcal{E}+\delta_a\left[\delta_b\left(\Sigma^b+\varepsilon^{bc}\Omega_c\right)\right].
\end{eqnarray}

Since $\mathcal{E}<0$, coupled with \eqref{norm20}, the $r=2$ horizon in the Schwarzschild background evolves into an isolated (\textit{resp.} dynamical) horizon provided that 

\begin{eqnarray}\label{norm21}
\delta_a\left(Z^a+\Sigma^a+\varepsilon^{ab}\Omega_b\right)=\mbox{(\textit{resp.} $>$)} 0.
\end{eqnarray}
Separately, we examine the two cases of \eqref{norm21}.

\subsubsection{Null case}

We start with the null criterion, with the vanishing of \eqref{norm21}, in harmonics, yielding

\begin{eqnarray}\label{norm23}
0=-\frac{\bar k^2}{r^2}\left[\mathcal{A}_V+2\left(\Sigma_V-\bar\Omega_V\right)\right],
\end{eqnarray}
which should hold for all $\bar k$.

By defining $Y_a=\delta_a\mathcal{E}$ and decomposing the equation \eqref{norm20} into harmonics, we write it down as the pair of equations

\begin{align}
Y_V=-\frac{\bar k^2}{r^2}\left(\Sigma_V-\bar\Omega_V\right);\quad\bar Y_V=0.\label{norm24}
\end{align}
Upon substituting the first equation of \eqref{norm24} into \eqref{norm23} gives

\begin{eqnarray}\label{norm25}
0=-\frac{\bar k^2}{r^2}\mathcal{A}_V+2Y_V.
\end{eqnarray}
Evoking the vanishing of the Greenberg vector, \eqref{norm24} is $r^2Y_V+\bar k^2\alpha_V=0$, so that \eqref{norm25} reduces to the condition

\begin{eqnarray}\label{norm28}
\mathcal{A}_V+2\alpha_V=0.
\end{eqnarray}
This is the required condition that preserves the null character of the null horizon $r=2$ under linear perturbation.

Were one to fix the frame by setting $\mathcal{A}^a=0$ (so that $n^a$ points along $\dot{u}^a$), we will have that the condition \eqref{norm28} now becomes $\alpha_V=0$. Otherwise, the horizon does not retain its null character. 

On the other hand, by \eqref{norm28}, the vanishing condition $\Sigma_V-\bar\Omega_V=0$ would lead to the nullity condition fixing the frame: $\mathcal{A}_V=0$. \\

\noindent\textbf{FOTH}: Let us consider the FOTH condition, $\dot{\chi}<0$ and $\hat{\chi}>0$. If one fixes the frame by setting $\mathcal{A}^a=0$, the FOTH condition trivially follows: Explicitly, the FOTH condition is the pair

\begin{align}
\mathcal{E}&<-\delta_a\alpha^a,\label{comb1}\\
\mathcal{E}&<\delta_aa^a+\delta_a\alpha^a,\label{comb2}
\end{align}
where we have used the fact that the Greenberg vector must vanish. So, again, how the black hole evolves under linear perturbation depends crucially on how the frame is choosen, as one would expect.

Again, $a^a$ must vanish for our MOTS to be a genuine surface. As the specification of the frame choice $\mathcal{A}^a=0$ gives $\alpha_V=0$, the FOTH condition is simply the requirement that $\mathcal{E}<0$, which always holds.

\subsubsection{Dynamical case}

Let us now consider the case where the $r=2m$ null horizon possibly evolves to acquire a different character. (Again for simplicity let us make the same frame choice here by setting $\mathcal{A}^a=0$.) Then, with all considerations of the vanishing Greenberg vector, as well as the necessity of the vanishing of $a^a$, the characterizing norm $z$ takes the form

\begin{align}
z=-12\frac{\bar k^2}{r}\mathcal{E}\left(\Sigma_V-\bar\Omega_V\right).
\end{align}

We note that from the field equations the vectors $\Sigma_a$ and $\Omega_a$ do not evolve independently \cite{cc1,cc2}, and so the presence of one ensures the non-vanishing of the other. Hence, we see that as long as there is a nonvanishing shear, $z\neq0$ for $\bar k^2\neq0$, and if there is no shear the horizon stays null as was seen in the previous case. 

We do expect that under the perturbation and in the presence of a nonzero shear the Schwarzschild event horizon becomes spacelike. This then imposes that

\begin{align}
\Sigma_V-\bar\Omega_V<0,
\end{align}
thereby imposing a constraint on the solution to the harmonically decomposed linearized system, when restricted to the horizon. We summarize the above discussion as follows:\\

\noindent\textit{Let $\tilde{M}$ be the spacetime obtained by linearly perturbing a Schwarzschild background. Then the $r=2$ horizon is null in $\tilde{M}$ if and only if the shear and vorticity 2-vectors obey $\Sigma_V-\bar{\Omega}_V=0$, but otherwise spacelike in $\tilde{M}$ with $\Sigma_V-\bar{\Omega}_V<0$.}\\

\noindent\textit{Note:}  A nondissipative perturbation is assumed here. The above conclusion drawn therefore may or may not be true in the presence of a nonzero $Q$ and $\Pi$, depending on their relative magnitudes.

\subsection{Spatially homogeneous background}

We consider a hypersurface orthogonal LRS II solution, which is spatially homogeneous, with metric

\begin{align}
ds^2=-dt^2+\bar a^2_1(t)dr^2+\bar a^2_2(t)d\bar{\varsigma}^2.
\end{align} 
(The 2-surface with metric $d\bar{\varsigma}^2$ is allowed different geometries including a toroidal one.) In this case we have that \(\mathcal{A}=\phi=0\), with the non-vanishing background kinematic quantities being the expansion and shear:

\begin{align}
\theta=\frac{\bar a_{1t}}{\bar a_1}+2\frac{\bar a_{2t}}{\bar a_2};\qquad \sigma=\frac{2}{3}\left(\frac{\bar a_{1t}}{\bar a_1}-\frac{\bar a_{2t}}{\bar a_2}\right),
\end{align} 
where the underscore `$t$' indicates partial differentiation. Hence,

\begin{eqnarray*}
\chi_0=\frac{2}{3}\theta-\sigma\qquad\mbox{and}\qquad\chi_1=\phi.
\end{eqnarray*}
Indeed, \(\dot{\chi}_1\) and \(\hat{\chi}_0\) are first order quantities so that \(\hat{\chi}_0^2=\hat{\chi}_0\hat{\chi}_1=0\). For simplicity, we specify to a conformally flat perturbation, and set $\Lambda=0$. We further restrict to the nondissipative case. This then gives the norm \(z\) as (we recall that this is always evaluated at $\chi=0$)

\begin{align}\label{norm15}
z&=F-2\dot{\chi}_0\dot{\chi}_1\nonumber\\
&=\frac{1}{3}\left(\rho-3p\right)\left[\left(\rho+p\right)-\delta_aZ^a\right]-\frac{4}{3}\rho\delta_a\alpha^a.
\end{align}

Also, the vanishing condition \eqref{norm2} becomes (again evaluating at $\chi=0$)

\begin{align}\label{norm16}
0=\dot{\chi}_0\delta_a\dot{\chi}&=-\frac{1}{3}\left(\rho+3p\right)\delta_a\left(\delta_b\mathcal{A}^b+\delta_b\alpha^b\right)\nonumber\\
&=\frac{\bar{k}^2}{3r^2}\left(\rho+3p\right)\left(\mathcal{A}_V+\alpha_V\right)\mathcal{Q}_a.
\end{align}
So, either we fix the equation of state $\rho+3p=0$ or $\mathcal{A}_V+\alpha_V=0$.

Now, for the former case this would require that $\rho$ (and consequently $p$) vanishes in the background to ensure the gauge invariance of $z$, in which case the horizon is always null. 

Let us consider the latter case, and consider the particular frame choice $\mathcal{A}^a=0$. This would impose that $\alpha_V=0$. (Note again that this condition is compatible with $\Sigma_V-\bar\Omega_V=0$ by virtue of the vanishing of the Greenberg vector.) Therefore, the characterizing function $z$ reduces to

\begin{align}
z=\frac{1}{3}\left(\rho-3p\right)\left(\rho+p\right).
\end{align}
Hence, assuming the weak energy condition, the character of a null horizon, evolving under linear perturbation, behaves according to the sign of $\rho-3p$: the horizon stays null, or becomes spacelike or timelike provided $\rho=3p,\rho<3p$ or $\rho>3p$. So, for example, with a linear equation of state (EoS) $p=w\rho$, the above conditions are the respective requirements $w=0, w>0$ and $w<0$. 

Of course, a choice different from $\mathcal{A}^a=0$ may be used to fix the frame. In this case, we will have $z$ assuming the form

\begin{align}
z=\frac{1}{3}\left[\left(\rho-3p\right)\left(\rho+p\right)+\frac{\bar k^2}{r}\left(5\rho-3p\right)\alpha_V\right].
\end{align}

We see that even with a linear EoS of the form $p=w\rho$, the signature of $z$ above acquires some complexity an carries a crucial dependence on sign of the component $\alpha_V$ (Note that $\alpha_V=\Sigma_V-\bar{\Omega}_V$). We may however still comment on some possible characters of the perturbed horizon (we are assuming a positive $\rho$). For example, it is not difficult to see that for the range

\begin{align}
-1\leq w\leq\frac{1}{3},\label{eqst1}
\end{align}
$\alpha_V>0$ precludes a spacelike or null character, allowing for only a timelike character. This indeed differs from the timelike characterization of the background as the extremes $w=-1$ and $w=1/3$ allow only for the null character. (See for example the references \cite{iben1,as1}.) If we look outside the bound \eqref{eqst1} in the case $\alpha_V>0$, for $w<-1$, the horizon character will depend on the magnitude of $\alpha_V$ at each $r$ in relation to each $\bar k^2$. (In the background, this is quite straightforward as outside the bound \eqref{eqst1}, the horizon is spacelike. Again, see the references \cite{iben1,as1}.) This will also be the case for the bound

\begin{align}
\frac{1}{3}< w<\frac{5}{3},\label{eqst2}
\end{align}
but for $w\geq5/3$, only the timelike character is possible.

Now consider the case for $\alpha_V<0$. Then, a timelike or null character is ruled out within the bound \eqref{eqst2}, allowing for only a spacelike character. For the bounds

\begin{align}
-1< w<\frac{1}{3}\qquad\mbox{and}\qquad w>\frac{5}{3},\label{eqst5}
\end{align}
the character again depends on the magnitude of $\alpha_V$ at each $r$ in relation to each $\bar k^2$.

At the extreme $w=1/3$ the horizon is spacelike and timelike for

\begin{align}
\frac{1}{3}< w\leq\frac{5}{3},\label{eqstz}
\end{align}
And for values $w<-1$, a timelike or null character is prohibited. (Notice that at the critical point $w=5/3$ the horizon signature is insensitive to the sign of $\alpha_V$.) 

We summarize the above results as follows:\\

\textit{Let $\tilde{M}$ be the spacetime obtained by linearly perturbing a spatially homogeneous background $M$ with EoS $p=w\rho$, and let $\mathcal{T}$ be a null horizon foliated by MOTS in $M$. Then for $\Sigma_V-\bar{\Omega}_V=0$, $\mathcal{T}$ is null, spacelike or timelike in $\tilde{M}$ provided $w=0, w>0$ or $w<0$, respectively. Otherwise, for $\Sigma_V-\bar{\Omega}_V\neq0$, see Table \ref{table1}. (Note that $\alpha_V=\Sigma_V-\bar{\Omega}_V$ on the MOTS.)} 

\begin{table}[h!]
\begin{center}
\begin{tabular}{ |p{2cm}||p{4cm}|p{2cm}|  }
 \hline
 \multicolumn{3}{|c|}{\textbf{Null horizon character under linear perturbation}} \\
 \hline
\quad\quad $\alpha_V$ & \qquad\qquad\qquad $w$  & \quad Character\\
 \hline
  \  &\  & \ \\
\  & \quad$(-1\leq w\leq\frac{1}{3})\cup (w\geq\frac{5}{3})$    &\quad Timelike\\
\quad $\alpha_V>0$ &\  & \ \\
\  &   \quad$(\frac{1}{3}< w<\frac{5}{3})\cup (w<-1)$  & \quad\qquad-- \\
\  &\  & \ \\
\hline
\  &\  & \ \\
\  &\quad$(-1< w<\frac{1}{3})\cup (w>\frac{5}{3})$ &\quad\qquad--\\
\  &\  & \ \\
\quad$\alpha_V<0$ &  \qquad\qquad$\frac{1}{3}<w\leq\frac{5}{3}$  & \quad Timelike\\
\  &\  & \ \\
\ &\qquad $(w\leq-1)\cup (w=\frac{1}{3})$  & \quad Spacelike \\
 \  &\  & \ \\
 \hline
\end{tabular}
\caption{The behavior of a null horizon evolving under linear perturbation. Assuming a linear equation of state $p=w\rho$, the above table shows the character of the horizon for different ranges of the equation of state parameter $w$, and the relationship with the sheet vector $\alpha^a$ decomposed harmonically. The `$-$' indicates a non-trivial relationship which is dependent on the magnitudes of $\alpha_V$ and $\rho$ as well as the value of the nonnegative integer $\bar k^2$.}
\label{table1}
 \end{center}
 \end{table}
 
 Let it be emphasized that the linear choice of the EoS is for demonstrable purposes for our particular application and there is no restriction here. Any EoS is in principle allowable with (possible) caveat being the the analysis acquiring a more complicated character.

We will now conclude this section with the following remark:

\begin{remark}[Remark 3.]
In a general inhomogeneous case, $\chi_1=0$ and we see that the associated norm is simplified:

\begin{align}
z=F,\nonumber
\end{align}
so that the characterization proceeds from \eqref{gip3}. However, the consideration of the $\delta_az=0$ criterion is now significantly more complicated. Relaxing this condition of course then gives a simple pointwise characterization via \eqref{gip3}. If one does not impose $\delta_az=0$ \textit{a priori}, this allows for the possibility of varying character across the horizon.  This is clearly a more involving case and this is worth considering for a subsequent work. However, as was just mentioned, while generally these cases are expected to pose problems, we see that in a general inhomogeneous background the condition provides a certain analytic simplicity, at least point-wise.
\end{remark}



\section{On the MOTS stabiliity operator}\label{sec5}


As have been seen from the previous section, the 2-vectors associated to the shear and vorticity play a crucial role in how a null horizon in a LRS background spacetime evolves under a linear perturbation. This particular combination of the shear and vorticity evolves simultaneously according to the field equations and have no independent evolution or propagation equations as was mentioned in the previous section. This relationship has implications for the self-adjointness of the MOTS stability operator as these same 2-vectors also control the self-adjointness of the operator. We will begin by introducing the MOTS stability operator after which we provide some commentary on this relationship. We will then delve into several properties of the operator in context of our current considerations as it relates to self-adjointness. The evolution, propagation and sheet equations that will be utilized here have been obtained in \cite{cc1,cc2} (with corrections in \cite{sem1,sem2} as stated in Section \ref{sec3}). So, especially the reference \cite{cc1} is implicit wherever these equations appear.

\subsection{The MOTS stability operator}

Given a MOTS $\mathcal{S}$ and a normal foliation by a collection $\mathcal{S}_v$, with the subscript `$v$' labelling the foliation, one may define the tangent vector to curves generating the deformation as

\begin{eqnarray*}
\partial_v=\bar\psi n^a,
\end{eqnarray*}
for some smooth function \(\bar\psi\) on \(\mathcal{S}\). Then, stability of $\mathcal{S}$ is captured via the variation of the expansion \(\chi\) along \(\partial_v\) (several discussions of this can be found in the original papers \cite{and1,and2}, as well as, for example, the works \cite{jara1,ib6} and references therein):

\begin{eqnarray*}
\bar\delta_{\bar\psi n}\chi=L_{\mathcal{S}}\bar\psi,
\end{eqnarray*}
where the operator $L_{\mathcal{S}}$ is 

\begin{eqnarray}\label{a1}
L_{\mathcal{S}}=-\delta^2+2\tilde s^a\delta_a+\bar F,
\end{eqnarray}
with the definition

\begin{eqnarray}\label{an}
\bar F=\frac{1}{2}\mathcal{R}_{\mathcal{S}}-\tilde s_a\tilde s^a+\delta_as^a-G_{ab}k^al^b,
\end{eqnarray}
(It is important to note that the scalar $\bar\psi$ completely characterizes the variation of $\mathcal{S}$.) The one-form $\tilde s_a=-N^c_{\ a}l_b\nabla_ck^b$ is the projection of the torsion of $k^a$ onto $\mathcal{S}$, $G_{ab}$ is the Einstein tensor and $\mathcal{R}_{\mathcal{S}}$ is the scalar curvature of $\mathcal{S}$.

$\mathcal{S}$ is said to be strictly or marginally stable if there exists a nonnegative (and not identically zero) $\bar{\psi}$ obeying respectively $L_{\mathcal{S}}\bar{\psi}>0$ or $L_{\mathcal{S}}\bar{\psi}=0$, and unstable otherwise. 

For a 1+1+2 decomposed spacetime, $\tilde s_a$ is explicitly

\begin{eqnarray}\label{tan3}
\tilde{s}_a=\Sigma_a-\varepsilon_{ab}\Omega^b,
\end{eqnarray}
And in the harmonic basis,
   
\begin{eqnarray}\label{tan4}
\tilde s_a=\underbrace{\left(\Sigma_V+\bar{\Omega}_V\right)}_\text{$\tilde{s}_V$}\mathcal{Q}_a+\underbrace{\left(\bar{\Sigma}_V-\Omega_V\right)}_\text{$\bar{\tilde{s}}_V$}\bar{\mathcal{Q}}_a.
\end{eqnarray}

In general the operator $L_{\mathcal{S}}$ is not self-adjoint, which results from the presence of the linear term $\tilde{s}^a\delta_a$ so that $L_{\mathcal{S}}$ is allowed to have complex eigenvalues. The imaginary part encodes information about rotation through the Komar angular momentum, obtained by integrating over the MOTS the inner product of $\tilde{s}_a$ and an axial Killing field.

Still, the principal eigenvalue, which we denote as $\bar{\lambda}$, is always real \cite{and1}. The stability of a MOTS $\mathcal{S}$ then reduces to the sign of $\bar{\lambda}$: $\mathcal{S}$ is strictly or marginally stable if $\bar{\lambda}>0$ or $\bar{\lambda}=0$, and unstable otherwise.

Here, we are interested in the implications of the characterization conditions obtained for the backgrounds in the previous sections, for the self-adjointness of the stability operator. Crucially, the dependence on the sign of $\alpha_V$ (or equivalently $\Sigma_V-\bar{\Omega}_V$). 

Now, it is known that the self-adjointness of the MOTS stability operator is guaranteed by the one-form $\tilde{s}_a$ being a gradient \cite{ib6}. Particular, if the 1-form $\tilde{s}_a$ is gradient, then the operator $L_{\mathcal{S}}$ is similar to a self-adjoint operator. (Of course, the case of vanishing $\tilde{s}_a$ is trivial, and so we consider this possibility $\tilde{s}_a\neq0$.) This self-adjointness in the case that $\tilde{s}_a$ is gradient can be seen here in a rather trivial way. To see this, let us consider the action of the operator $L_{\mathcal{S}}$ on a first order scalar $\psi^{(1)}=\psi_S\mathcal{Q}$:

\begin{eqnarray}\label{st5}
L_{\mathcal{S}}y^{(1)}=\left[\left(-\delta^2+\bar F\right)\mathcal{Q}+\frac{1}{r}\tilde{s}_V\right]\psi_S.
\end{eqnarray}
(The normalization $\mathcal{Q}_a\mathcal{Q}^a=1$ has been assumed for simplicity.) Since if the  $\tilde{s}_a$ is gradient it has a vanishing divergence, it is then clear that this requires the vanishing of $\tilde{s}_V$, i.e.

\begin{align}
\Sigma_V+\bar{\Omega}_V=0.\label{selfa1}
\end{align}
Thus, the eigenvalue problem for the MOTS stability operator simply reduces to the following eigenvalue problem for \(\mathcal{Q}\):

\begin{eqnarray}\label{st6}
-\delta^2\mathcal{Q}+F\mathcal{Q}=\lambda \mathcal{Q}.
\end{eqnarray}
That is, the problem reduces to the eigenvalue problem for the self-adjoint operator $L'_{\mathcal{S}}\mathcal{Q}=\left(-\delta^2+F\right)\mathcal{Q}=\lambda\mathcal{Q}$ so that

\begin{eqnarray}\label{st6}
\lambda_{\bar{k}^2}=\frac{\bar{k}^2}{r}+\bar{F},
\end{eqnarray}
i.e. the eigenvalues are parametrized by $\bar{k}^2$. Thus, strict stability is always guaranteed for 

\begin{align}
\bar{F}=\frac{1}{2}\left(\mathcal{R}_{\mathcal{S}}-R\right)-\bar{\tilde{s}}_V^2-\left(p+Q\right)+2\Lambda>0.
\end{align}
(recall that $R$ is the four dimensional scalar curvature.) 

It will later be seen that $\bar{\tilde{s}}_V$ is necessarily zero on the MOTS in the case that $\tilde{s}_V=0$, i.e. $\tilde{s}_a$ is a gradient. So, in this case for $p+Q\leq0$ and $\Lambda\geq0$, whenever $\mathcal{R}_{\mathcal{S}}>R$, the operator has no negative eigenvalue. The principal eigenvalue and the MOTS is strictly stable. This is indeed the case, say, for vacuum solutions undergoing a shear-free and a nondissipative perturbation where one will have $\tilde{s}_V$ vanishing.

On the other hand, consider an irradiating ($Q=0$) spacetime with a positive scalar curvature, a vanishing $\Lambda$ and a nonnegative pressure. Then, for a sufficiently large $p$ it is possible that $\bar{F}<0$, so that the principal eigenvalue is always negative and hence the MOTS will be unstable.

\begin{remark}[Remark 4.]
Actually, in the linear regime any perturbation to the MOTS scalar curvature $\mathcal{R}_{\mathcal{S}}$ is encoded in first order scalars. Particularly, if the scalars $\mathcal{E},\Pi$, and $\Lambda$ (if $\Lambda$ is included as a perturbation variable) are nonvanishing in the background, then $\mathcal{R}_{\mathcal{S}}$ is not perturbed. However, nonlinear effects are present under perturbation of the background as 

\begin{align}
\mathcal{R}_{\mathcal{S}}=\ ^0\mathcal{R}_{\mathcal{S}}+\mathcal{Z},
\end{align}
where $\mathcal{Z}=\Sigma_{ab}\Sigma^{ab}-\zeta_{ab}\zeta^{ab}$ is just twice the inner product of the projected shears of $k^a$ and $l^a$. So, it is clear that once nonlinear effects are factored in any analysis, we do expect the behavior of the eigenvalues to be modified.
\end{remark} 

Now, the commutation relations of the dot and hat derivatives for an arbitrary 2-vector $\Psi_a$ (imposing the vanishing of the Greenberg vector and taken up to linear order)

\begin{align}
\hat{\dot{\Psi}}_a-\dot{\hat{\Psi}}_a=-\mathcal{A}\dot{\Psi}_a+\left(\frac{1}{3}\theta+\sigma\right)\hat{\Psi}_a+\mathcal{H}\varepsilon_{ab}\Psi^b.
\end{align}
Harmonically decomposing the above relation and applying to the one-form $\tilde{s}_a$ while setting the component $\tilde{s}_V=0$ says that

\begin{align}
\hat{\dot{\bar{\tilde{s}}}}_V-\dot{\hat{\bar{\tilde{s}}}}_V&=-\mathcal{A}\dot{\bar{\tilde{s}}}_V+\left(\frac{1}{3}\theta+\sigma\right)\hat{\bar{\tilde{s}}}_V,\label{st10}\\
\mathcal{H}\bar{\tilde{s}}_V&=0.\label{st11}
\end{align}

The above is meant to suggest that for a magnetized ($\mathcal{H}\neq0$) linear perturbation $\tilde{M}$ of a LRS background spacetime $M$, if the 1-form $\tilde{s}_a$ is gradient in which case the MOTS stability operator is self-adjoint (remember this is for the class of MOTS under consideration), then $\tilde{s}_a$ necessarily vanishes on the MOTS. But as it will be seen shortly, the quantities $\bar{\tilde{s}}_V$ and $\mathcal{H}$ are proportional so that in the linear regime on any MOTS it is true that both $\bar{\tilde{s}}_V$ and $\mathcal{H}$ vanish. More specifically, $\mathcal{H}$ must be first order. 

\subsection{Comments on relations to the characterization of null horizons}

Let us now specify our discussions to the backgrounds of the previous section.

\subsubsection{Schwarzschild background}

To begin with, consider the case of the Schwarzschild background. We will establish the following statement (C1) and the corresponding implication (C2):

\begin{enumerate}

\item[(C1).] \textit{For a linear perturbation $\tilde{M}$ of a Schwarzschild background, the 1-form $\tilde{s}_a$ is gradient (in which case the MOTS stability operator is self-adjoint) if and only if $\tilde{s}_a$ is identically zero; $\implies$}

\item[(C2).] \textit{For a linear perturbation $\tilde{M}$ of a Schwarzschild background, if the one-form $\tilde{s}_a\neq0$ and the vorticity 2-vector has a nonzero contribution from the odd sector, the $r=2$ null horizon necessarily evolves to a spacelike character.}
\end{enumerate}

\noindent\textit{Note:} As was mentioned in the previous section, away from the null case the horizon can only evolve to the spacelike case with the even parity component of $\Sigma_a+\varepsilon_{ab}\Omega^a$ being negative. (A timelike character would imply that timelike curves do enter the trapped region, which is of course ruled out.) Thus, the second statement above, (C2), follows from the fact that $\tilde{s}_V,\bar{\Omega}_V\neq0$ ensures that $\Sigma_V-\bar{\Omega}_V\neq0$.

Let us apply the tensor $\varepsilon^{ab}n^c$ to the Ricci identities for the unit field $n^a$ to get

\begin{align}
\delta_a\left(\Omega^a+\varepsilon^{ab}\Sigma_b\right)=\left(2\mathcal{A}-\phi\right)\Omega+\mathcal{H},\label{st14}
\end{align}
and upon noting $\Omega$ should be zero on the MOTS, the above reduces to 

\begin{align}
-\frac{\bar{k}^2}{r}\bar{\tilde{s}}_V\mathcal{Q}=\mathcal{H},
\end{align}
so that $\mathcal{H}$ is a first order quantity that vanishes on the background, i.e. $\mathcal{H}$ is decomposed into the harmonic $\mathcal{Q}$ basis.

Then, by \eqref{st11} it follows that $\mathcal{H}=\bar{\tilde{s}}_V=0$. It therefore follows that $\tilde{s}_V=0\iff \tilde{s}_a=0$, i.e. $\tilde{s}_a$ is gradient if and only if $\tilde{s}_a$ vanishes identically. 

Finally, the linearized evolution and propagation equations of $\mathcal{H}$ are respectively given by (these are respectively the $u^a$ and $n^a$ components of the magnetic Weyl divergence equation \cite{el4})

\begin{align}
\dot{\mathcal{H}}&=-3\mathcal{E}\xi-\varepsilon_{ab}\delta^a\mathcal{E}^b,\label{st12}\\
\hat{\mathcal{H}}&=-3\mathcal{E}\Omega-3\phi\mathcal{H}-\delta_a\mathcal{H}^a,\label{st13}
\end{align}
which translates to, by the vanishing of $\mathcal{H}$ as well as noting that $\Omega=\xi=0$ on the MOTS,

\begin{align}
\mathcal{H}_V=0;\quad \bar{\mathcal{E}}_V=0.\label{st21}
\end{align}
This is to say that, necessarily, on the MOTS one requires that $\mathcal{H}^a$ is solenoidal (divergence-free) and that $\mathcal{E}^a$ has no odd parity contribution on the MOTS.

While there appears to be freedom in the quantities $\mathcal{E}_V$ and $\bar{\mathcal{H}}_V$, they are actually constrained by a coupling of the sheet expansion $\phi$ and the 1-form $\tilde{s}_a$ (see the discussions of the next subsection). For example, for a vanishing $\tilde{s}_a$ in the case of a nondissipative perturbation, it follows that one requires (again, see the next subsection for more details on this)

\begin{align}
\mathcal{E}_V+\bar{\mathcal{H}}_V=0.\nonumber
\end{align}

Of course, we do not rule out self-adjointness of the operator in the case of a spacelike character. This is in principle possible depending on the nature of the perturbation. This case will however impose that $\Sigma_V<0$. 

On the other hand the null character does not necessarily imply the self-adjointness of the operator. The self-adjointness due to vanishing of $\tilde{s}_a$, in this case, would imply (and is implied by) $\Sigma_V=\bar{\Omega}_V=0$.

\subsubsection{Hypersurface orthogonal and spatially homogeneous background}

We now consider the  spatially homogeneous case of the previous section, and stick with a nondissipative perturbation. In this case the equations corresponding to \eqref{st12}, \eqref{st13} and \eqref{st14} are respectively

\begin{align}
\dot{\mathcal{H}}&=-\frac{3}{2}\left(\frac{2}{3}\theta-\sigma\right)\mathcal{H}-3\mathcal{E}\xi-\varepsilon_{ab}\delta^a\mathcal{E}^b,\label{st15}\\
\hat{\mathcal{H}}&=-3\left(\mathcal{E}+\rho+p\right)\Omega-3\phi\mathcal{H}-\delta_a\mathcal{H}^a,\label{st16}\\
\mathcal{H}&=\delta_a\left(\Omega^a+\varepsilon^{ab}\Sigma_b\right)+3\xi\sigma.\label{st17}
\end{align}
The vanishing of $\Omega,\xi=0$ and $\mathcal{H}$ then implies the result $\mathcal{H}_V,\bar{\mathcal{E}}_V,\bar{\tilde{s}}_V=0$ follows as in the Schwarzschild case. For this reason, the statement (C1) of the previous subsection also holds here, although for obvious reasons the statement (C2) does not hold.

\begin{remark}[Remark 5.]
We do expect these three vanishing properties to be generic to the MOTS we are considering in the sense that they are independent of the particular background one works with. It will be quite interesting to see how much these vanishing conditions further constrains the MOTS in full generality. This is however beyond the scope of the current work. 
\end{remark} 

In the self-adjoint case with a vanishing $\tilde{s}_a$, obviously either one of the components $\Sigma_V$ or $\bar{\Omega}_V$, paired with the parameter ranges suffice for the characterization of the horizon.

On the other hand, if one is to impose the condition that $|\Sigma_V|\neq|\bar{\Omega}_V|$, this of course ensures that $\tilde{s}_a$ is nonvanishing, in which case one may then check which conditions this further impose on the horizon character. However, this does not in by itself rule out self-adjointness of the operator. In principle, if one can expand a nonvanishing $\tilde{s}_V$ in the harmonic $\mathcal{Q}$ basis, then the operator can still be brought to a self-adjoint form. In the forthcoming subsection we will see a particular class of perturbations where this may be possible.

\subsection{More on the MOTS and stability operator}

We obtain some additional results required to be satisfied on the MOTS, given properties derived from the one-form $\tilde{s}_a$. As we shall see, this will provide some interesting insights into properties of the 1-form $\tilde{s}_a$. 

The magnetic Weyl tensor is written down as a constraint to the field equation \cite{el4}:

\begin{align}
H_{ab}=\varepsilon_{cd\langle a}D^c\sigma_{b\rangle}^{\ d}-2\dot{u}_{\langle a}\omega_{b\rangle}-D_{\langle a}\omega_{b\rangle}.\label{mw1}
\end{align}
One may interpret the above as the magnetic Weyl tensor measuring the degree of distortion of vorticity. 

Taking the $n^a$ component of \eqref{mw1} produces the constraint equation, up to linear order,

\begin{align}
-\delta_a\left(\frac{2}{3}\theta-\sigma\right)&=-2\delta^b\Sigma_{ab}-\phi\tilde{s}_a\nonumber\\
&-2\varepsilon_{ab}\left[\mathcal{H}^b+\left(\delta^b+2\mathcal{A}^b\right)\Omega\right]\nonumber\\
&+\xi\left(\Omega_a-3\varepsilon_{ab}\Sigma^b\right).\label{hd1}
\end{align}
By setting $\Omega=\xi=0$ we have

\begin{align}
-\delta_a\left(\frac{2}{3}\theta-\sigma\right)=-2\left(\delta^b\Sigma_{ab}+\varepsilon_{ab}\mathcal{H}^b\right)-\phi\tilde{s}_a.\label{hd2}
\end{align}
Also, fully projecting the Ricci identities for $n^a$ taken up to linear order and imposing the MOTS condition and the vanishing of $\xi$ and $\Omega$ as required we obtain

\begin{align}
-\delta_a\phi=2\left(\mathcal{E}_a-\delta^b\zeta_{ab}\right)-\Pi_a-\phi\tilde{s}_a,\label{mw13}
\end{align}
and upon combining with \eqref{hd2} and \eqref{mw13} we have

\begin{align}
-\cancel{\delta}_a\chi=2\left[\mathcal{E}_a-\delta^b\left(\Sigma_{ab}+\zeta_{ab}\right)-\varepsilon_{ab}\mathcal{H}^b\right]-\phi\tilde{s}_a,\label{mw13n}
\end{align}
where we have written $\cancel{\delta}_a=\delta_a-\tilde{s}_a$. One may then use the Codazzi equations to establish the following

\begin{align}
-Q_a+Z_{ab}k^b=2\left(\mathcal{E}_a-\varepsilon_{ab}\mathcal{H}^b\right)-\phi\tilde{s}_a,\label{mw14}
\end{align}
where $Z_{ab}=N^{\ c}_aC_{cbde}k^dl^e$, with $C_{abcd}$ being the spacetime Weyl tensor.

We will now utilize a curvature property of the MOTS to establish that the second term on the left hand side of \eqref{mw14} is of nonlinear order on the MOTS. With this we will have established a relationship between the Weyl and heat flux 2-vectors and the rotation 1-form. 

The curvature 2-form on the MOTS, i.e. the curvature of $\tilde{s}_a$ -- which we denote by $\tilde{\Omega}_{ab}$ -- contracted along the null direction $k^a$ can be shown to obey the relation

\begin{align}
k^b\tilde{\Omega}_{ba}=-N^{\ c}_aZ_{cb}k^b,\nonumber
\end{align}
so that

\begin{align}
k^b\tilde{\Omega}_{ba}=-\left[2\left(\mathcal{E}_a-\varepsilon_{ab}\mathcal{H}^b\right)-\phi\tilde{s}_a+Q_a\right].\label{mw15}
\end{align}

Explicitly, the left hand side of \eqref{mw15} can be reduced to

\begin{align}
k^b\tilde{\Omega}_{ba}&=-\frac{1}{2}\Sigma_b\delta_ak^b\nonumber\\
&=-\frac{1}{2}\left(\Sigma_{ab}+\zeta_{ab}\right)\Sigma^b,\nonumber
\end{align}
which we ignore as it is of second order. Thus,

\begin{align}
\phi\tilde{s}_a=Q_a+2\left(\mathcal{E}_a-\varepsilon_{ab}\mathcal{H}^b\right).\label{mw16}
\end{align}

Furthermore, imposing the MOTS condition on \eqref{coom2} we have 

\begin{align}
\cancel{\delta}_a\phi=0,\nonumber
\end{align}
so that by \eqref{mw13} it follows

\begin{align}
\Pi_a=2\left(\mathcal{E}_a-\delta^b\zeta_{ab}\right).\label{mw17}
\end{align}

Clearly, away from the case of a vanishing $\phi$, the following statement is concluded:\\

\noindent\textit{The MOTS stability operator in a conformally flat and nondissipative (this imposes that $\zeta_{ab}$ is divergence-free) linearly perturbed inhomogeneous LRS background is self-adjoint}.\\

So for example, the above will be true for a nondissipative perturbation of a Lemaitre-Tolman-Bondi type solutions.

Notice that if the tensor $\delta_an_b$ is pure trace, dissipation on the MOTS is generated entirely by contributions from the Weyl tensor.

On the other hand, without restricting to the nondissipative case, using the vanishing of $\bar{\mathcal{E}}_V,\mathcal{H}_V$ and $\bar{\tilde{s}}_V$ we have the following pair of equations from \eqref{mw16}.

\begin{align}
\bar{Q}_V&=0,\label{mw18}\\
\phi\tilde{s}_V&=Q_V+2\left(\mathcal{E}_V+\bar{\mathcal{H}}_V\right).\label{mw19}
\end{align}

In any case, we can draw some immediate conclusions from both \eqref{mw16} and \eqref{mw17} about the generation of rotation on a horizon:

\begin{itemize}

\item \noindent\textit{In a conformally flat linearly perturbed inhomogeneous LRS background, rotation of a horizon foliated by MOTS can be entirely sourced by the heat flux 2-vector along the MOTS. This then implies that self-adjointness of the MOTS stability operator in this particular case may be obtained by switching off any heat flux contribution along the MOTS.} 

\item \noindent\textit{In a nondissipative linearly perturbed inhomogeneous LRS background, rotation of a horizon foliated by MOTS can be sourced by the 2-vector components of the Weyl curvature tensor. This is to say that in this case the contributing sources are tidal forces acting along the MOTS.} 

\end{itemize}

\begin{remark}[Remark 6.] We remark here that the form of \eqref{mw19} suggests that it is possible, in principle, to have a nonvanishing $\tilde{s}_V$ -- i.e. a nonvanishing $\tilde{s}_a$ -- even in the case of a conformally flat perturbation with a vanishing heat flux and vanishing $\phi$. Thus, in the first bulleted point above we were quite cautious with semantics when we used \textit{``may be''} 
 when switching off the heat flux contribution  along the MOTS to obtain self-adjointness of the stability operator. Only when $\phi$ is nonzero in the perturbed spacetime that this is true, at least in our current context.\end{remark}

Finally, let us describe a simple case where $\tilde{s}_V$ is non-vanishing but the stability operator may take a self-adjoint form. More specifically, we may be able to express $\tilde{s}_V$ in terms of the $\mathcal{Q}$ basis. We shall consider a static background, i.e. the shear and expansion scalars vanish on the background.

We note that for any 2-tensor $\Psi_{ab}$ one has as the divergence

\begin{align}
\delta^b\Psi_{ab}=\frac{1}{r}\left(\bar{k}^2-2\right)\left(-\bar{\Psi}_T\mathcal{Q}_a+\Psi_T\bar{\mathcal{Q}}_a\right).
\end{align} 
Thus, one can harmonically decompose \eqref{hd2} to obtain the pair of constraint equations

\begin{align}
0&=\frac{1}{r}\left(\bar{k}^2-2\right)\Sigma_T+\mathcal{H}_V,\label{mw10}\\
0&=\frac{1}{r}\left(\bar{k}^2-2\right)\bar{\Sigma}_T+\bar{\mathcal{H}}_V\nonumber\\
&+\frac{1}{2r}\left(\frac{2}{3}\theta_S-\sigma_S\right)-\phi\tilde{s}_V.\label{mw11}
\end{align}

Indeed, from the first equation \eqref{mw10} it follows that the tensorial part $\Sigma_T$, for $\bar{k}^2\neq2$, vanishes by virtue of the vanishing of the quantity $\mathcal{H}_V$. Furthermore, by imposing the MOTS condition $\chi=0$, the second equation \eqref{mw11} can be recast as

\begin{align}
0=\frac{1}{r}\left(\bar{k}^2-2\right)\bar{\Sigma}_T+\bar{\mathcal{H}}_V+\phi\left(\frac{1}{2r\mathcal{Q}}-\tilde{s}_V\right).\label{mw25}
\end{align}
Therefore, as long as the MOTS is nonminimal in the perturbed spacetime (recall that by nonminimal it is meant that $\phi$ and $(2/3)\theta-\sigma$ cannot vanish simultaneously on the MOTS), for a perturbation with no contributions from the odd vector harmonics part of $\mathcal{H}_a$ and the shear 2-tensor $\Sigma_{ab}$, the 1-form component $\tilde{s}_V$ can effectively be cast as 

\begin{align}
\tilde{s}_V=\left(\frac{1}{2r\mathcal{Q}^2}\right)\mathcal{Q}.\label{mw26}
\end{align}
(Of course $\tilde{s}_V$ is nonzero and so $\tilde{s}_a$ is not a gradient.) This then allows us to write down the eigenvalues of the operator as

\begin{eqnarray}\label{st6}
\lambda_{\bar{k}^2}=\frac{\bar{k}^2}{r}+\frac{1}{2r^2\mathcal{Q}^2}+\bar{F}.
\end{eqnarray}

Consequently the positivity of the modification term means that $\bar{F}>0$ again ensures nonnegativity of the eigenvalues.


\section{Summary and outlook}\label{sec6}


\noindent\textit{Summary} -- In this work we have considered the behavior of a null horizon foliated by MOTS in a LRS background spacetime subjected to linear perturbations. The problem is fixed by the gauge invariant requirement of the characterizing function. More precisely, we set out to characterize the evolution of a MOTS in a linearized spacetime admitting a 1+1+2 decomposition, which lies in the slice orthogonal to the $u^a$. The gauge invariance of the chacterizing function introduced is possible if and only if a certain function vanishes in the background. This vanishing condition is equivalent to the condition for a horizon in the background to be null, thus phrasing the problem in terms of the evolutionary behavior of a null horizon in a LRS background subjected to linear perturbations. In other words -- and emphasis is placed on this -- this characterization fails in the case of a non-null horizon in the background. While this work has been dedicated to the case of linear perturbations, it was demonstrated that the characterizing function is form invariant even when extending to non-linear regimes, for a certain frame choice or provided that the perturbed spacetime has a vanishing sheet expansion.

We focused the application of our characterization to non-dissipative perturbations. As was described in the introduction, perturbing a seed metric generally affects the dynamics of an embedded horizon, and that this is the case of the null event horizon in the Schwarzschild spacetime. The presence of shear ensures a spacelike transition of the null horizon. We have assumed a harmonic decomposition of first order scalars $\Psi$ and 2-vectors $\Psi_a$, where the notations $\Psi_V$ and $\bar{\Psi}_V$ are used respectively for the components of $\Psi_a$ along the even and odd vector harmonics. It is established here that this spacelike transition requires that a particular relationship between the shear and the vorticity 2-vectors holds, specifically, $\Sigma_V-\bar{\Omega}_V<0$.  In order for the causal preservation of the horizon character, $\Sigma_V-\bar{\Omega}_V=0$. The result was obtained for the particular consideration of a non-dissipative perturbation where the scalars, tensors and vector quantities associated to the heat flux and anisotropic pressure are vanish.

We also considered the case of a spatially homogeneous background solution and wrote down the form of the horizon characterizing function. In the Schwarzschild case, any transition from a null character is necessarily spacelike as the horizon encloses a black hole (note that we are assuming the MOTS condition is preserved). However, for the spatially homogeneous case null horizons do not enclose black holes, and therefore the causal character could transition to a timelike one under the perturbation. Imposing a linear equation of state, for a particular frame choice, i.e., fixing the spatial unit vector to align with the acceleration of the temporal unit vector (which coincides with the vanishing of the difference between the 2-vector components, i.e., $\Sigma_V-\bar{\Omega}_V=0$), the sign of the equation of state parameter $w$ alone characterizes the evolution: null if $w=0$, spacelike if $w>0$, and timelike if $w<0$. Without this frame choice, the evolution of the horizon is a bit more involved. We nonetheless present a partial characterization of the horizon dynamics in yhje presence of perturbations. In particular, where the difference between these vector components is strictly negative or strictly positive, we find the ranges of the equation of state parameter where the induced metric on the horizon has absolute sign, i.e., the causal character is timelike or spacelike. However, there are ranges of the parameter for which the causal character is determined by a complex relationship between the 2-vectors, the energy density and the eigenvalues of MOTS Laplacian, which we could not analytically determine.

Finally, we considered the relationship between the shear and vorticity 2-vectors which play a crucial role in the characterization of the horizon, to the MOTS stability operator. These 2-vectors also specify the 1-form $\tilde{s}_a$ which controls the self-adjointness of the operator, and points to an obvious connection to the characterization we have introduced. We established that the component $\bar{\tilde{s}}_V$ vanishes for our particular consideration in this work, and that $\tilde{s}_V=\Sigma_V+\bar{\Omega}_V$. This implied that for our consideration the 1-form $\tilde{s}_a$ is a gradient (so that the MOTS stability operator is self-adjoint) if and only if $\tilde{s}_a$ vanishes identically. As a consequence, in the case of perturbation of a Schwarzschild background, it follows that when the stability operator is not self-adjoint and $\bar{\Omega}_V\neq0$, the event horizon necessarily transitions to a spacelike character. In the case of a spatially homogeneous background, in the self-adjoint case with a vanishing $\tilde{s}_a$, it is clear that only one of the components $\Sigma_V$ or $\bar{\Omega}_V$ is needed for the characterization. 

Several additional results were also obtained, restricting the form of 2-tensors and 2-vectors on the MOTS. For example, it was established that (and this appears independent of the choice of background) the components of the even and odd vector harmonics, respectively, of the magnetic and electric Weyl 2-vectors necessarily vanish on the MOTS. In fact, it was demonstrated that the bi-implication $\bar{\tilde{s}}_V=0\iff\mathcal{H}=0$ follows, which is due to the fact that $\mathcal{H}$ is necessarily a first order quantity. Furthermore, the component of the odd tensor harmonics of the shear 2-tensor also vanishes on the MOTS. 

As the 1-form $\tilde{s}_a$ is tied to rotation generation on the horizon, we provide a clear picture, at least in the linear regime, of the precise source of rotation on the horizon: for a non-dissipative linear perturbation of a LRS background spacetime, rotation is entirely sourced by contributions from the Weyl 2-vectors on the MOTS. If one includes dissipation, then rotation gains a contribution from the heat flux 2-vector. Actually, there could be dissipation from the anisotropic 2-vector and no contribution from the heat flux 2-vector, and it remains true that horizon rotation will be entirely sourced by Weyl contributions. It then follows that for a conformally flat and non-dissipative perturbation of an inhomogeneous LRS background spacetime, the stability operator is necessarily self-adjoint.

We also presented a class of perturbations for which the 1-form $\tilde{s}_a$ is non-vanishing but one may nonetheless expand the component $\bar{\tilde{s}}_V$ in the harmonic basis so that the MOTS stability operator can be brought to a self-adjoint form.

\ \\
\noindent\textit{Outlook} -- There are several interesting directions that could be explored. In this work, we have restricted ourselves to first order perturbations. What happens when non-linearities are incorporated into the analysis of the horizon evolution? Naturally the problem becomes a bit (or a lot) more complicated as seen from the characterizing function $z$ in (27). For example, there will be additional contributions from the non-linearities to the scalar curvature of the MOTS, so that large distortions are introduced on the MOTS. These distortions are not present at first order as was discussed earlier. Modification to the eigenvalue in the self-adjoint case will present in the form of the inclusion of dissipation and the cosmological constant $\Lambda$ (if $\Lambda$ is considered a perturbation `variable'). Any non-linearity introduced is an extra contribution to the principal eigenvalue through the scalar curvature of the MOTS. 

Another example is the following. At every stage of our analysis we have imposed the MOTS condition $\chi=0$, meaning that the built-in assumption here is that a background horizon, under perturbations of its ambient background spacetime, evolves to a horizon foliated by  MOTS. However, it is quite possible that under the perturbation the horizon is no longer sliced into MOTS. This is the case of the Weyl-distorted Schwarzschild geometry discussed in the introduction, where by the tuning up the distortion parameter the $r=2$ horizon is no longer a MOTS \cite{pil1}. This degeneracy may be a feature of non-linearities under the perturbation and so might not happen in the linear perturbative case considered in this work. However, as mentioned in the second point above, extending our analysis to non-linear perturbations, it would suffice as a more general consideration, not to impose the MOTS condition and analyze the non-linear equations.


\section{Acknowledgement}


AS and SK research is supported by the Basic Science Research Program through the National Research Foundation of Korea (NRF) funded by the Ministry of education (grant numbers) (NRF-2022R1I1A1A01053784) and (NRF-2021R1A2C1005748). PKSD acknowledges support through the First Rand Bank, South Africa. AS would also like to thank the IBS Center for Theoretical Physics of the Universe, Daejeon, South Korea, for its hospitality, where a part of this work was carried out.

\end{document}